\documentclass[12pt,preprint]{aastex}
\shorttitle{Inner-shell absorption lines}
\shortauthors{Behar \& Netzer}
 
\def\micron{\ifmmode \mu{\rm m} \else $\mu$m\fi}
\def\kms{\ifmmode {\rm km\,s}^{-1} \else km\,s$^{-1}$\fi}
\def\Hubble{\ifmmode {\rm km\,s}^{-1}\,{\rm Mpc}^{-1}
        \else km\,s$^{-1}$\,Mpc$^{-1}$\fi}
\def\ergsec{\ifmmode {\rm ergs\;s}^{-1} \else ergs s$^{-1}$\fi}
\def\ergscm{\ifmmode {\rm ergs\,s}^{-1}\,{\rm cm}^{-2}
          \else ergs\,s$^{-1}$\,cm$^{-2}$\fi}
\def\ergscmA{\ifmmode {\rm ergs\,s}^{-1}\,{\rm cm}^{-2}\,{\rm \AA}^{-1}
          \else ergs\,s$^{-1}$\,cm$^{-2}$\,\AA$^{-1}$\fi}
\def\ergscmHz{\ifmmode {\rm ergs\,s}^{-1}\,{\rm cm}^{-2}\,{\rm Hz}^{-1}
          \else ergs\,s$^{-1}$\,cm$^{-2}$\,Hz$^{-1}$\fi}
\def\Msun{\ifmmode M_{\odot} \else $M_{\odot}$\fi}
\def\Lsun{\ifmmode L_{\odot} \else $L_{\odot}$\fi}
\def\qo{\ifmmode q_{0} \else $q_{0}$\fi}
\def\Ho{\ifmmode H_{0} \else $H_{0}$\fi}
\def\ho{\ifmmode h_{0} \else $h_{0}$\fi}
\def\qo{\ifmmode q_{0} \else $q_{0}$\fi}
\def\ao{\ifmmode a_{0} \else $a_{0}$\fi}
\def\to{\ifmmode t_{0} \else $t_{0}$\fi}
\def\Halpha{\ifmmode {\rm H}\alpha \else H$\alpha$\fi}
\def\Hbeta{\ifmmode {\rm H}\beta \else H$\beta$\fi}
\def\hb{\ifmmode {\rm H}\beta \else H$\beta$\fi}
\def\Hgamma{\ifmmode {\rm H}\gamma \else H$\gamma$\fi}
\def\Hdelta{\ifmmode {\rm H}\delta \else H$\delta$\fi}
\def\Lya{\ifmmode {\rm Ly}\alpha \else Ly$\alpha$\fi}
\def\Lyb{\ifmmode {\rm Ly}\beta \else Ly$\beta$\fi}
\def\hi{\ifmmode \mbox{{\rm H}\,{\sc i}} \else H\,{\sc i}\fi}

\def\ciii{\ifmmode {\rm C}\,{\sc iii} \else C\,{\sc iii}\fi}   
\def\civ{C\,{\sc iv}\,$\lambda1549$}

\def\oiii{[O\,{\sc iii}]\,$\lambda5007$}

\def\o5007{[O\,{\sc iii}]\,$\lambda5007$}
\def  \kms         {\hbox{km s$^{-1}$}}          

\def  \La          {\ifmmode {\rm Ly}\alpha \else Ly$\alpha$\fi}
\def  \Ka          {\ifmmode {\rm K}\alpha \else K$\alpha$\fi}
\def  \Lb          {\ifmmode {\rm L}\beta \else L$\beta$\fi}
\def  \Ha          {\ifmmode {\rm H}\alpha \else H$\alpha$\fi}
\def  \Hb          {\ifmmode {\rm H}\beta \else H$\beta$\fi}
\def  \Pa          {\ifmmode {\rm P}\alpha \else P$\alpha$\fi}
\def  \CIIIb       {\ifmmode {\rm C}\,{\sc iii]}\,\lambda1909
                     \else C\,{\sc iii]}\,$\lambda1909$\fi}
\def  \CIV         {\ifmmode {\rm C}\,{\sc iv}\,\lambda1549
                     \else C\,{\sc iv}\,$\lambda1549$\fi}
\def  \MgII         {\ifmmode {\rm Mg}\,{\sc ii}\,\lambda2798
                     \else Mg\,{\sc ii}\,$\lambda2798$\fi}
\def  \OVI         {\ifmmode {\rm O}\,{\sc vi}\,\lambda1035
x
                     \else O\,{\sc vi}\,$\lambda1035$\fi}

\begin{document}
 
\title{The largest black holes and the most luminous galaxies}
\author{Hagai Netzer \altaffilmark{1}}
 
\altaffiltext{1}{School of Physics and Astronomy, Raymond and Beverly Sackler
Faculty of Exact Sciences, Tel-Aviv University, Tel-Aviv 69978, Israel.}
 
\received{October 2002}
\revised{}
\accepted{}
 
\shorttitle{Largest black holes}
\shortauthors{Netzer, H.}
 
\begin{abstract}                    
The empirical relationship between the broad line region size and the
source luminosity in active galactic nuclei (AGNs) is used to obtain 
black holes (BH) masses for a large number of quasars in three samples.
The largests BH masses found exceed  $10^{10}$ \Msun\
and are correlated, almost linearly, with the source luminosity. Such BH
masses, when converted to galactic bulge mass and luminosity, indicate
masses in excess of $10^{13}$ \Msun\ and ${\sigma_*}$ in excess of 700 km/sec.
Such massive galaxies have never been observed. The largest BHs reside,
almost exclusively,  in 
high redshift quasars. This, and the deduced BH masses, 
suggest  that several scenarios of BH and galaxy formation  are inconsistent
with  the observations. Either the observed size-L relationship in low
luminosity  AGNs does not extend to  very high luminosity or else the 
$M_{\rm BH} - M_{\rm bulge} - \sigma_*$ correlations observed in the local
universe do not reflect the relations of those quantities at the epoch of
galaxy formation.

\end{abstract}

\section{Introduction}
Recent progress in reverberation mapping of active galacic nuclei (AGNs) 
allowed the first meaningful correlation between the
 broad line region (BLR) size ($R_{\rm BLR}$) and the black hole (BH)
mass in more than 30 objects. This provided a simple way to
calculate  BH masses for   a large number of
sources and  resulted in a flood of papers on this topic. Some papers
(e.g. Vestergaard 2002, hereafter V02; 
McLure and Jarvis 2002) 
investigated, in great detail, the wavelength dependence  of the 
$R_{\rm BLR}-L-M$ relationship and provided
useful ways for adopting the  method to other wavelength bands.
 This opens the way for the study of  BH masses in large
samples of high luminosity high-z quasars.

All the new BH mass estimates are based on a single relationship obtained for a single
sample of 34 AGNs for which BLR sizes are available from  decade long
reverberation mapping  campaigns. More than half the sample was observed at
the Wise observatory over a  period of about 12 years (Kaspi et al. 2000,
hereafter K00). Other objects have been monitored in  other
observatories and in several ``AGN watch'' campaigns (Netzer \&
Peterson 1997; Peterson  2001). The main
findings are a significant  $R_{\rm BLR}- \lambda L_{\lambda} (5100)$ 
relationship  ($L_{\lambda} (5100)$ is the monochromatic luminosity at
5100\AA) and the confirmation that 
 the BLR gas is in virial motion (e.g. Peterson and Wandel
2000). These, plus the (model dependent) conversion 
of  the   observed  full-width at half maximum (FWHM) of various
emission lines into 3-D gas velocities, are sufficient to derive the mass of
the central BH. 

This letter discusses the mass of the largest BH in the universe; those  
 found in the centers of the most luminous quasars. It follows the works of
Laor (1998; 2001), McLure \& Jarvis (2002), Woo \& Urry (2002)  and others who 
used such methods for obtaining BH masses beyond the original K00 sample. The
paper addresses also the
Shields et al. (2002) new results and extends the mass estimates to much
larger quasar samples. Section 2  presents new mass calculations  for a
large number of  sources and  \S3  illustrates the new correlations
found. Section 4 discusses the new results in light of the available
information on the largest, most luminous elliptical galaxies and the epochs of
quasars and galaxy formation.

\section{The largest BH}

\subsection{BH Mass measurements}
 New mass estimates have been obtained  for a large number of AGNs using
 the $R_{\rm BLR}-L$
relationship obtained from  the K00 sample; the only sample available for
such  calibration. This relationship is
given, schematically, by 
\begin{equation}
   R_{\rm BLR} = c_1 L_{\lambda}^{\gamma}
\end{equation}
 which results in the
following mass estimate:
\begin{equation}
M_{\rm BH} = c_2   L_{\lambda}^{\gamma} [FWHM]^2 \,\,.
\end{equation}
Here $c_1$ and $c_2$ are  constants that include the flux
normalization and 
various assumptions about the velocity field in the BLR. The slope
 ${\gamma}$ is derived from the reverberation campaigns results and is in the
range 0.5-0.7 (see below).  The expression in eqn. 2  can be  used to derived 
``single epoch'' masses   that combine the constants $\gamma$ and $c_2$ 
 with  observed  FWHM  of
certain emission lines in  {\it individual} objects. The method has been
described in various  papers including K00, V02, and McLure \& Jarvis (2002).
Its  more useful applications are based on 
measured $\lambda L_{\lambda}$(5100) 
and FWHM(\hb)  for low redshift sources (the quantities used by
K00) and  the combination of $\lambda L_{\lambda}$(1350) and FWHM(\civ) 
for high redshift objects. 
V02 has looked into  the inter-calibration 
of  the two and supplied
the  expressions that are used in this work except for a small correction
in the value of $c_2$ that was introduced  to adjust her constants to the
cosmology assumed here:  
  $H_0=70$ \Hubble, 
$\Omega_m=0.3$ and $\Omega_\Lambda=0.7$.
McLure \& Jarvis (2002) provided similar expressions
for $MgII \, \lambda 2798$ which are not used in this work.

\subsection{The sample}
Three AGN samples have been used in this work:
1)  the LBQS sample (Forster et al. 2000 and references
therein),  2) A sample of 104 high redshift high luminosity quasars with 
ground-based  spectrophotometry 
($L_{\lambda}(1350)$) and good FWHM(\civ) measurements, and 3) the new 
$L_{\lambda}(5100)$ and FWHM(\hb) listed by Shields et al (2002). Many of the sources
in the second sample are UM quasars and the raw data   can be found in
MacAlpine and Feldman 1982, Baldwin, Wampler \& Gaskell (1987) and Baldwin (1977).

Forster et al. (2000) supplied monochromatic luminosities and FWHMs for many
emission lines  in about 1000  LBQS quasars.
Since many of the sources have been observed through relatively small aperture, and
under poor  weather conditions, it was decided to use the Bj magnitudes that
are much more accurate. This  follows Green et al. (2001)
who studied the Baldwin relationship in this sample and  obtained
monochromatic luminosities  using the same method.
All fluxes have been corrected for galactic reddening using the  Green et al.
procedure. A major assumption here, and in Green et al. (2001),  is that 
the observed continuum can be described by a single   
$L_{\nu} \propto \nu^{-\alpha}$ power-law with $\alpha=0.5$. This approximation
neglects the possible dependence of $\alpha$ on source luminosity which may
affect the $L-M$ relationship (see \S 4). Forster et al. (2001) provided several different measurements
of FWHM(\civ) with and without the narrow line component. The
``single'' component fit was used and the ``broad only'' fit was checked to
verify that the results are not sensitive to this choice. A handful of sources
with FWHM(\hb)$<1,000$ \kms\ or with FWHM(\civ)$>20,000$ \kms\ were removed from 
the sample since those were considered unreliable or affected too much by the narrow 
emission line.
  As for the second \civ\ sample, no
galactic reddening was applied and the same assumption about $L_{\nu}$ was
used. In this case there is no significant  dependence on $\alpha$ since the
original papers quote the observed flux at around rest wavelength of
1450\AA.

The above  samples are optically selected and suffer from various  selection effects. This is of no
real consequence to the main goal of the paper which is to derive the mass of
the largest known BHs. It may affect, however,  the derived $M-L$ correlations
(\S4).  

\section{The $L-M$ relationship for high luminosity AGNs}
BH masses have been calculated using equation 2 and the normalizations  derived
by K00 and V02 adjusted to the cosmology chosen here. The
determination of the slope $\gamma$ is crucial for the present work and will
be discussed prior to presentation of the new results.

We start from the original K00 sample to which we apply two statistical
methods for finding $\gamma$: 
the Akritas \& Bershady  (1996) BCES
estimator (for which we only consider the BCES bisector) and the {\it fitexy} method
described in Press et al. (1992). The merits of the different  methods have
been discussed, extensively, in several papers and will not be repeated here.
Our experience shows that  that the differences between the slopes obtained by
the different methods  are larger than the formal
uncertainties on the slopes of each method.
The K00 sample adjusted to the new cosmology gives $\gamma=0.58 \pm 0.12$ for
the BCES bisector estimator and $\gamma=0.68 \pm 0.03$ for the {\it fitexy} method.
The two are formally consistent with each other and $\gamma ({\rm BCES})$ was
adopted here.

Since the purpose of this work is to extrapolate to very large $L$, we also
experimented with removing the lowest luminosity objects from the sample.
Removing the three objects with $\lambda L_{\lambda}(5100)< 10^{43}$ \ergsec\
resulted in 
$\gamma ({\rm BCES})=0.71 \pm 0.21$ and  $\gamma (fitexy)=0.69 \pm 0.03$.
Removing
the seven objects with $\lambda L_{\lambda}(5100)< 10^{43.7}$ \ergsec\ resulted
in 
$\gamma ({\rm BCES})=0.58 \pm 0.19$ and  $\gamma (fitexy)=0.74 \pm 0.04$.
All these results suggest that the two methods are consistent with each
other and the slope cannot be determined to an accuracy better than about 0.15.
The value adopted for illustrating the results of 
 this work is the smaller one found for the entire
K00, $\gamma=0.58$. The implications
for the case of larger or smaller $\gamma$ are discussed in \S4.

Shields et al. (2002) suggested the use of the ``physically motivated'' value
of $\gamma=0.5$. The strongest argument for using this value is the suggestion
by Netzer \& Laor (1993) that the outer boundary of the BLR is determined by
the dust sublimation radius which is similar to the measure $R_{\rm BLR}$ to
within a factor $\sim 2$. There are several problems in applying this idea
to the present mass determination.
 First,  the ``reverberation
radius'' is determined by the responsivity of \hb\ to changes in the ionizing
luminosity, $L_{\rm ion}$, which is smaller than the bolometric luminosity
that determines the dust sublimation radius. In addition,  
 $\gamma=0.5$ means  the same BLR ionization parameter for low
luminosity Seyferts and the highest luminosity quasars. This has never been
shown to be the case in large QSO samples. Thus, more work is required to
justify this theoretical value of $\gamma$.

The masses computed with the $\gamma=0.58$  slope are
presented in Fig. 1. The diagram contains data for 505 QSOs with \civ\
measurements and 219 source with \hb\ measuremets. The luminosity range is
roughly 
$\lambda L_{\lambda}(1350) = 10^{44-47.5}$ \ergsec. Also shown  is the
best regression line (see below) and the  mass range of $\pm \sigma_M$ around
the median calculated in luminosity bins of 0.3 dex. The largest BHs  are found in
sources with $z>2$ with $M_{\rm BH} \simeq 10^{10.2}$ \Msun\ (15 
  with mass exceeding $10^{10}$ \Msun). Using
$\gamma=0.68$ (the slope found with the $fitexy$ method),
 raise this number to about $10^{10.4}$ \Msun\ (62 with mass exceeding $10^{10}$ \Msun).
 
%
%
\centerline{\includegraphics[width=15cm]{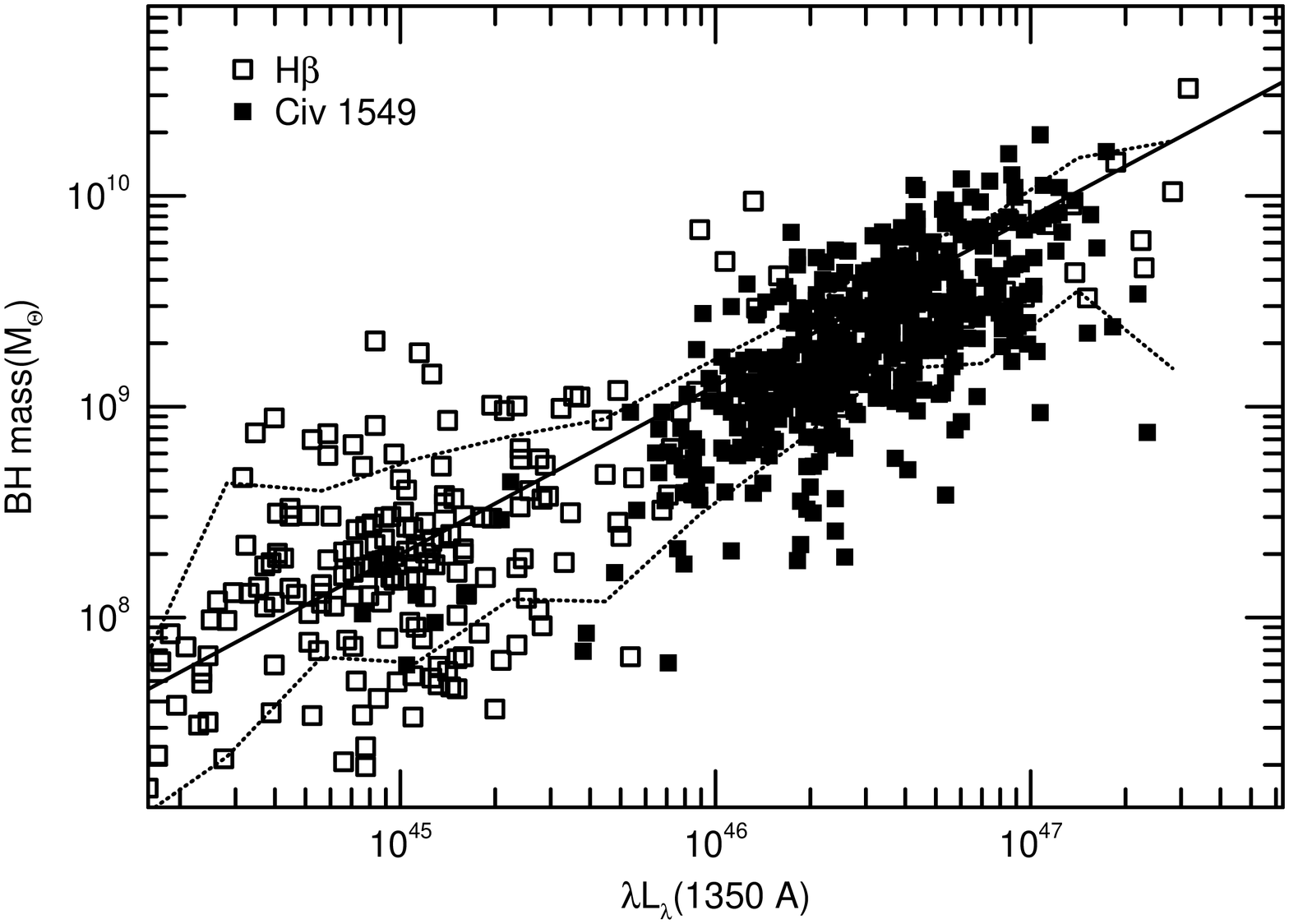}} 
\figcaption {
Black hole mass as a function of $\lambda
L_{\lambda}$(1350) for the quasar samples described in the text assuming
$\gamma=0.58$. Open symbols represent mass obtained from the \hb\ line and full
symbols masses obtained using \civ.  The two \civ\
samples completely overlap in properties and were not
given different symbols. The dashed lines represent the $\pm \sigma_M$ range
around the median  and the straight line the $M \propto L^{0.8}$
relationship.
\label{fig1_label}
 }
\centerline{}
\centerline{} 
Fig. 2 shows $M_{\rm BH}$ as
a function of redshift for the same sample under the same assumptions.

\centerline{\includegraphics[width=15cm]{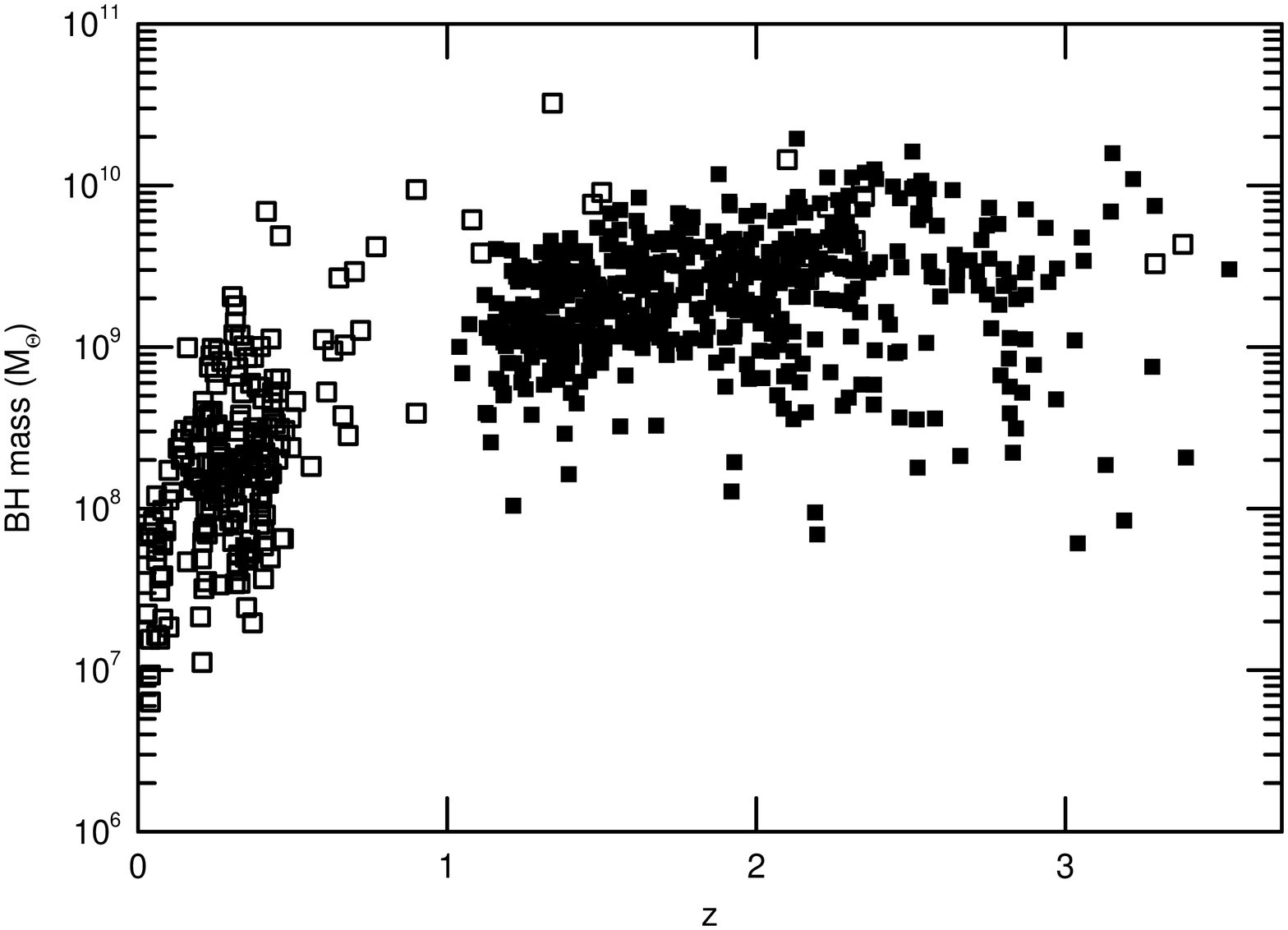}}  
\figcaption
{
    $M_{\rm BH}$ vs. redshift for the sample in figure 1 with the same
symbols. \label{fig2_label}
 } 
\centerline{}
\centerline{}   

The data in Fig. 1 suggest a simple linear dependence of the form
 $M_{\rm BH} \propto L^{\beta}$. This has been tested by
performing a linear regression analysis using the same two methods described
earlier.
The procedure
used for calculating the errors is the following:  For $L_{1350}$, the
assumption is of a constant error of 0.15 dex representing the measurement
uncertainty, the extrapolation in wavelength and the typical range in
luminosity due to continuum variability. This number does not affect the
resulting  slope $\beta$ in any significant way. As for the
mass, this was done using standard  error propagation 
combining all errors due to the uncertainties in $L$ and in FWHM (line width
uncertainties are  given in Forster et al. 2000). The combined error for this
case is typically 0.15-0.25 dex. No uncertainties are listed for FWHM(\civ) in 
the second quasar sample and for FWHM(\hb) in  Shields et al. (2002).
 A uniform error of 0.2
dex in  $M_{\rm BH}$  was assumed in those cases. The errors are relatively
large and are   expressed in logarithmic form (i.e.
0.5(log(x+dx)$-$log(x$-$dx), see Lyons, 1991).

Table  1 lists
several slopes obtained by the two methods for our standard case of $\gamma=0.58$  and
for $\gamma=0.68$, the slope obtained by the $fitexy$ method.
Given the various biases and unknowns, it is reasonable to assume that the
real  uncertainty in $\beta$ is at least as large as $\pm 0.15$.
With this uncertainty, the slopes of the 
  \civ\ sample and the entire sample are barely consistent with each other
 and the slopes of
the \hb\ sample and the entire sample are indistinguishable. The scatter in
slope is probably due to the very different luminosity range of the \civ\ and
the \hb\ samples.
 A second approach that was tries assumed a uniform  uncertainty in
$M_{\rm BH}$ of 0.3 dex for all objects. This gave very similar results.
  The overall conclusion is that for luminous AGNs, 
 $M_{\rm BH} \propto L^{0.9 \pm 0.15}$. 

\begin{deluxetable}{lcccc}
\tablewidth{0pt}
\tablecaption{Regression analysis results for $M_{\rm BH} =a
L_{1350}^{\beta}$ }
 \tablehead{
  \colhead{method} &
  \colhead{$\gamma$} &
  \colhead{sample   } &
  \colhead{$\beta$ }  &
   \colhead{$a$}
}
\startdata    
BCES Bisector & 0.58 &all \civ\ & 1.12$\pm 0.05$ & -42.9 \\
$fitexy$ &  0.58 & all \civ\ & 1.13$\pm 0.04$ & -43.1 \\
BCES Bisector & 0.58 & all \hb\ & 0.83$\pm 0.03$ & -29.2 \\
$fitexy$ & 0.58 &all \hb\  &0.80$\pm 0.02$ & -27.6 \\
BCES Bisector & 0.58 & all objects& 0.80$\pm 0.02$ & -27.7 \\
$fitexy$ & 0.58 & all objects & 0.78$\pm 0.01$ & -26.9 \\
BCES Bisector & 0.68 & all objects& 0.90$\pm 0.02$ & -32.2 \\
$fitexy$ & 0.68 & all objects & 0.89$\pm 0.01$ & -31.9 \\
\enddata  
\end{deluxetable}

The $M_{\rm BH}-L$ correlation found here  is very different from the one
found in K00. The reason is probably the incompleteness of the small
K00 sample which resulted in a biased sampling of  FWHM(\hb) vs. $\lambda
L_{\lambda}(5100)$  not representing  the parent population.
Indeed,  K00 found FWHM(\hb)$ \propto L^{-0.27}$ while in the samples under
 study the correlation is much flatter.
 The FWHM--luminosity dependence in various samples will be addressed in a
separate paper (Corbett et al. 2003).

The tight $M_{\rm BH}-L$ relationship enables the study of the Eddington
ratio, $L/L_{\rm Edd}$, in these samples.  The observed $M-L$
relationship suggests a very weak, if any  dependence of $L/L_{\rm Edd}$
on $L$ or on $M_{\rm BH}$. This impression is confirmed by a formal statistical
analysis. Since the  results are
 marginal, they will not be presented here. Another important issue is
 the mean  
 $L/L_{\rm Edd}$. This depends on the distribution in this property as
well as on the exact conversion from $\lambda L_{\lambda}$
 to bolometric luminosity and the value of $\gamma$.
Assuming  first $\gamma=0.58$ and 
$L_{\rm Edd} =9 \lambda L_{\lambda}(5100)$, as in K00, gives 
a median $L/L_{\rm Edd}$ of 0.53. 
The composite spectra published recently by
Telfer et al. (2002)  suggest a different conversion with 
$L_{\rm Edd} \simeq 5 \lambda L_{\lambda}(5100)$. This translates to 
a median of 0.28.
The above  values are transformed  to  0.33
 and 0.18 for the case of $\gamma=0.68$. In both cases the distribution
wis wide, covering about a factor 10 in $L_{\rm Edd}$.
 Thus, the choice of $\gamma=0.58$ results in a
large mean $L/L_{\rm Edd}$ and a large number of sources with 
super Eddington luminosities.
As explained by Woo and Urry (2002), the implications to
the derived $M_{\rm BH}-L$ relation are very importance (see \S4).

\section{Discussion: the largest BHs and the most luminous galaxies}
The new results presented here suggest that the largest BHs are situated in
the most luminous quasars that are, typically, the highest redshift sources. At
the extreme end of the distribution we find BH masses of order 
$5 \times 10^{10}$ \Msun\ if $\gamma=0.7$, 
 and
$1.1 \times 10^{10}$ \Msun\ if $\gamma=0.5$. 
This is  greater than obtained so far in large samples.
 The recent work by 
Shields et al. (2002) aimed at the calibration of the
the \oiii\ line width as a bulge mass   estimator.
 The method is based on the close agreement between
FWHM(\oiii) and the stellar velocity dispersion
$\sigma_*$ at low luminosity and the  K00 mass estimates at higher luminosity.
Using this method and  $\gamma=0.5$ (their Table 2) they find one object with
$M_{\rm BH}$ exceeding $10^{10}$ \Msun\ and several  others approaching this mass.
As shown in Fig. 1,  the \civ\
samples includes  many more sources with such large  masses.

Before addressing the cosmological 
 consequences we note the various factors influencing the   
 $M-L$ relationship and  likely reasons for overestimating $M_{\rm BH}$.

\begin{enumerate} 
\item
The K00 sample covers a limited luminosity range and all mass estimates
corresponding to  $\lambda L_{\lambda}(1350) >10^{46}$ are necessarily
obtained by extrapolation. Since this is {\it the only} sample
available so far, there is no independent way to verify the largest
masses until successful reverberation
mappings are obtained for higher luminosity AGN.  
Moreover, as explained in \S2,  the slope of the $R_{\rm BLR}$-$L$ relationship
is uncertain. The slope chosen here ($\gamma=0.58$) is close to the middle of
the range. Its increase to 0.7 will increase the mass at the high luminosity
end by a factor of about 2.5. 
 \item
The largest new mass estimates are based on the measured
$\lambda L_{\lambda}(1350)$ which is scaled to the K00 luminosity assuming the
same spectral energy distribution (SED) for high and low luminosity AGNs. This
assumption has never been tested in large quasar samples.
The data for such test  are already
available (Telfer et al. 2002) but the results are not yet known. Intrinsic
reddening, in the quasar host galaxy, is another potential complication
related to the inter-calibration of optical and UV luminosities.   
\item
The FWHM(\civ) may not reflect the virial motion of the BLR gas in high 
luminosity quasars.
\item
The samples used here  suffer from various selection effects. This  
influence only slightly the largest derived masses   but can
affect much more the $L-M_{\rm BH}$ correlation. For example, magnitude limited
samples may not include the less luminous quasars, those with the smallest
$L/L_{\rm Edd}$. This results in a false impression of a very strong $M-L$
correlation. Woo and Urry (2002) investigated this idea in great detail and concluded
that all strong $M-L$ correlations obtained so far suffer from such a selection effect.
\end{enumerate}

The main conclusion of this work is  that the
largest BH masses are found in the highest luminosity quasars. The masses
of such BHs can reach
the extreme values of $10^{10.3-10.6}$ \Msun, depending on the value of $\gamma$.
Using recent conversions to host galaxy properties 
  one finds 
$M_{\rm bulge} \sim 10^{13.1-13.4}$ \Msun\ (Kormendy \& Gebhardt 2001),
$M_{\rm B,bulge} \sim -25$ mag (Kormendy 2001) and
$\sigma_*$ exceeding 800 km/sec (Tremaine et al. 2002).
   Such galaxies
have never been observed and are not predicted to exist by  standard galaxy
formation theories.

In principle,  this is still consistent with the
observations since the sources with the largest $M_{\rm BH}$ are the most
luminous ones  and will completely out-shine any  host galaxy.
Thus, there is no direct way to rule out the existence of such galaxies.
However, the 
theoretical implication are in conflict with recent
ideas that the largest galaxies attain their mass through a series of
mergers, a process that operates continuously to redshift 2 or smaller. A
similar difficulty is found for the BH growth since those same theories (e.g.
Haehnelt and Kauffman 2000; Yu and Tremaine 2002) assume that galactic
nuclei BHs increase their mass up to redshifts smaller than 2 by the same
series of mergers (or, perhaps, only through large mergers). Thus, the largest
BHs are predicted to be associated with the most massive galaxies at $z<2$, in
conflict with the data in Fig. 2.
It is clear that active BH with $M_{\rm BH}>10^{10}$ \Msun\ 
 are not found in the local universe.
It is also clear  that dormant BHs of this mass, or the galaxies with extreme properties
that are supposed to host such BHs, have never been found.
The whereabout of the huge BH formed at $z \simeq 3$  is thus
unknown.

 A more plausible
suggestion is that some or  all the conversion factors used to obtain the
galactic   mass, magnitude and $\sigma_*$ from the BH mass,  that are based
on measurements in the local universe, cannot be extrapolated to high
luminosity high redshift objects. Perhaps they are only valid  at $z<2$,
after galaxies and nuclear BHs have accumulated most of their mass.
If correct, this would mean that some ``normal looking'' galaxies contain extremely
massive BHs. A similar
 suggestion by Laor (2001) involves a  dependence of
$M_{\rm BH}/M_{\rm bulge}$ on the BH mass or  the absolute magnitude of the
host galaxy.

To conclude, either the measurements of BH masses presented here for the
  most luminous quasars are grossly overestimated,
 because of the reasons described above,  or else the relationships between BH masses
and various properties of their host galaxies  at
high $z$ are very different from those measured in the local universe.
 A second conclusion, which is less certain because of various
selection effects, is that for AGNs,  $M_{\rm BH} \propto L^{0.9 \pm 0.15}$.

\acknowledgements
The work described in this paper is based primarily on
a  decade-long AGN monitoring 
project at the Wise observatory. I am grateful to many of my colleagues and students that 
helped in making this into a very successful project. Special thanks go to Dan Maoz, Shai Kaspi
and Ohad Shemmer who led various parts of the project and without whom it would have been
impossible to bring it to completion. Useful discussions with Ari Laor are
gratefully acknowledged.
 This work is supported by the Israel Science Foundation grant 545/00.

\end{document}